%% file: DT_calibration.tex
\documentclass[conference]{IEEEtran}
\usepackage[top=0.7in, bottom=0.96in, left=0.65in, right=0.65in]{geometry} 
\input{input.tex}

\usepackage{graphicx}
\usepackage{epstopdf}
\usepackage{enumerate}
\usepackage{float}
\usepackage{makeidx}
\usepackage{cleveref}
\usepackage{url}
\usepackage{subcaption}
\usepackage{booktabs}
\usepackage{makecell}
\usepackage{caption}
\usepackage{adjustbox}
\usepackage{threeparttable}

\DeclareMathOperator*{\argsort}{arg\,sort}

\begin{document}
\bstctlcite{IEEEexample:BSTcontrol}

\title{Wireless Digital Twin Calibration: \\ Refining DFT-Domain Channel Information}

\author{
    \IEEEauthorblockN{Hao Luo\IEEEauthorrefmark{1}, Saeed R. Khosravirad\IEEEauthorrefmark{2}, and Ahmed Alkhateeb\IEEEauthorrefmark{1}}
    \IEEEauthorblockA{\IEEEauthorrefmark{1}School of Electrical, Computer, and Energy Engineering, Arizona State University \ \{h.luo, alkhateeb\}@asu.edu}
    \IEEEauthorblockA{\IEEEauthorrefmark{2}Nokia Bell Laboratories \ \{saeed.khosravirad\}@nokia-bell-labs.com}
}

\maketitle

\begin{abstract}	
    Wireless digital twins can be leveraged to provide site-specific synthetic channel information through precise physical modeling and signal propagation simulations. This can help reduce the overhead of channel state information (CSI) acquisition, particularly needed for large-scale MIMO systems. For high-quality digital twin channels, the classical approach is to increase the digital twin fidelity via more accurate modeling of the environment, propagation, and hardware. This, however, comes with high computational cost, making it unsuitable for real-time applications. In this paper, we propose a new framework that, instead of calibrating the digital twin model itself, calibrates the DFT-domain channel information to reduce the gap between the low-fidelity digital twin and its high-fidelity counterpart or the real world. This allows systems to leverage a low-complexity digital twin for generating real-time channel information without compromising quality. To evaluate the effectiveness of the proposed approach, we adopt codebook-based CSI feedback as a case study, where refined synthetic channel information is used to identify the most relevant DFT codewords for each user. Simulation results demonstrate the effectiveness of the proposed digital twin calibration approach in achieving high CSI acquisition accuracy while reducing the computational overhead of the digital twin. This paves the way for realizing digital twin assisted wireless systems.
\end{abstract}

\section{Introduction} 
    Current and future wireless communication networks rely heavily on scaling MIMO systems at the infrastructure and user devices. Employing larger antenna arrays can lead to potential gains in spatial multiplexing and array gains. Fully realizing these gains, however, requires acquiring channel state information (CSI) for downlink precoding design. In frequency division duplex systems, downlink CSI acquisition typically consists of three steps~\cite{3GPP}: (i) The base station (BS) sends pilot signals, and users estimate their channels based on the received pilot signals. (ii) Users compress and quantize their estimated channels using a predefined codebook and report the corresponding codeword indices and coefficients back to the BS. (iii) The BS reconstructs the CSI from the reported feedback and designs the downlink precoding vectors accordingly. This process incurs significant overhead, especially in large-scale MIMO systems, due to the need for extensive pilot transmission and feedback. Therefore, reducing the overhead of downlink CSI acquisition is crucial for enabling efficient communication in large-scale MIMO systems.
    
    Wireless digital twins~\cite{Alkhateeb2023} have emerged with the potential to benefit large-scale antenna systems. Through multi-modal sensing and measurement techniques, digital twins can replicate real-world communication environments using 3D geometric models, electromagnetic (EM) material properties, and hardware representations. Moreover, signal propagation simulations, such as ray tracing, can be used generate synthetic channel information for a given transmitter–receiver pair. This channel information can then serve as prior knowledge for CSI acquisition and reduce the overhead. Nonetheless, digital twins present their own set of challenges: (i) The digital twin may not be perfectly modeled, which leads to inaccurate channel information. (ii) Ray tracing simulations can be time-consuming, making them unsuitable for real-time applications. (iii) The communication environment is often dynamic, so digital twins need to adapt to changing conditions. Therefore, further research efforts are needed to fully realize the potential of digital twins in wireless communication systems.

    Prior work has explored how digital twins can assist real-time physical layer operations of wireless communication systems. For example, in~\cite{Alikhani2025}, a wireless digital twin is used to generate synthetic channel information for assisting downlink CSI acquisition. This type of application, however, requires not only accurate modeling but also fast simulations to support real-time operation. A potential solution is to use low-complexity ray tracing algorithms to accelerate the simulation process, though this often compromises the accuracy of the resulting channel information. Motivated by the inherent trade-off between the accuracy and complexity, in this paper, we study a lightweight approach to calibrate a low-complexity digital twin to refine its synthetic channel information. Notably, wireless digital twin calibration is a novel research area, and the calibration may be applied at different points in the processing chain. For instance, the twin itself could be calibrated if it is modular and parameterized (e.g., EM properties could be adjusted based on real-world measurements~\cite{Jiang2025}). Alternatively, one can calibrate its output, such as the synthetic channel information, which is the specific focus of this paper.

    \begin{figure*}[t]
        \centering
        \includegraphics[width=0.975\textwidth]{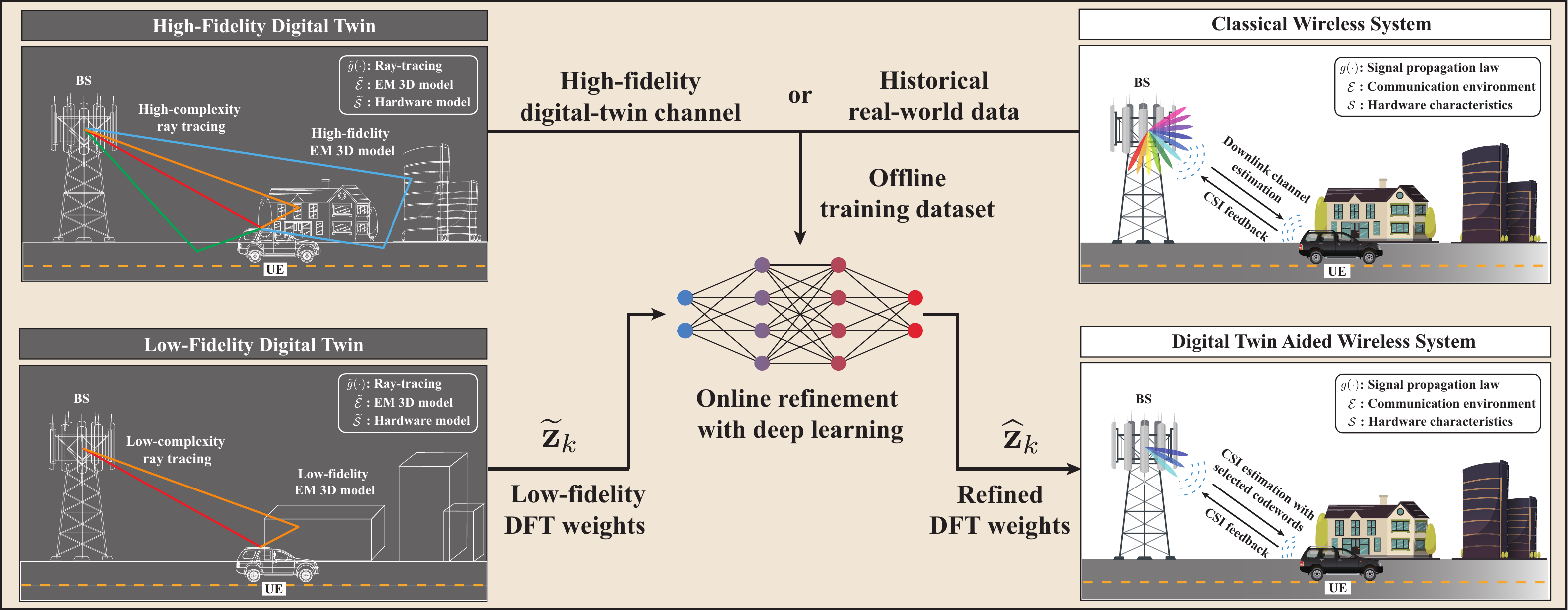}
        \caption{This figure illustrates the proposed digital twin calibration framework and its use case of codebook-based CSI feedback. The BS generates synthetic channel information for a user using a low-fidelity digital twin. The synthetic channel information is then refined using a lightweight deep learning model. The refined channel information is used to select the most relevant DFT codewords for that user. The user receives pilot signals corresponding to these selected codewords and estimates their coefficients, which are reported back to the BS for downlink precoder design.}
        \label{fig:system_model}
    \end{figure*}

    In this work, we propose a novel calibration approach that refines the synthetic channel information generated by a low-complexity digital twin. The calibration is implemented using a lightweight deep learning model, which can be efficiently trained via offline supervised learning. The contributions of this paper are summarized as follows:
    \begin{itemize}
        \item We propose a digital twin calibration approach that leverages a lightweight deep learning model to refine the synthetic channel information, which is represented by the discrete Fourier transform (DFT) weights of the user channel. The calibration model can be trained with offline supervised learning, using either high-fidelity digital twin data or historical CSI feedback from existing systems.
        \item We demonstrate the effectiveness of the proposed calibration approach with a case study of codebook-based CSI feedback. The refined synthetic channel information is used to identify the most relevant codewords, i.e., DFT beams, for each user. This enables more efficient downlink CSI feedback and reduces pilot signal overhead.
    \end{itemize}
    Simulation results highlight the effectiveness of our proposed approach in improving CSI acquisition accuracy while simultaneously reducing computational overhead of the digital twin. Furthermore, the framework enables a low-complexity digital twin to generate synthetic channel information which, through calibration, can achieve performance levels approaching those of high-fidelity digital twins. This capability brings us closer to the vision of real-time digital twins~\cite{Alkhateeb2023}. Finally, it is worth noting that the proposed methodology is not limited to codebook-based CSI feedback but can also be applied to beam management for mmWave frequency bands~\cite{Giordani2018}.

\section{System Model}
    As shown in \figref{fig:system_model}, we consider a MIMO system with $K$ single-antenna users served by a BS equipped with $N$ antennas. The BS is assisted by a digital twin that provides prior knowledge of the user channel, which is used to reduce the overhead of downlink CSI acquisition in the real world. Next, we present the system model, including the signal model, channel model, and digital twin model.

    \subsection{Signal and Channel Models}
        The BS precodes the transmitted symbol for the $k^{\text{th}}$ user, denoted as $s_k \in \bbC$, using a precoding vector $\bff_k \in \bbC^{N \times 1}$. The received signal at the $k^{\text{th}}$ user is given by
        \begin{equation}
            y_k = \bh_k^H \bff_k s_k + \sum_{l \neq k} \bh_k^H \bff_l s_l + n_k,
        \end{equation}
        where $\bh_k \in \bbC^{N \times 1}$ is the channel between the BS and the $k^{\text{th}}$ user, and $n_k \sim \cN(0, \sigma^2)$ is the receive noise at the $k^{\text{th}}$ user. A geometric channel model is considered, where the channel between the BS and the $k^{\text{th}}$ user is represented as a sum of $L_k$ multi-path components, given by
        \begin{equation}
            \bh_k = \sum_{l=1}^{L_k} \alpha_{k,l} \ba(\phi_{k,l}, \theta_{k, l}),
        \end{equation}
        where $\alpha_{k,l}$ is the complex gain, and $\ba(\phi_{k,l}, \theta_{k, l})$ is the array response vector at the BS, which depends on the azimuth and elevation angles of departure (AoD) $\phi_{k,l}$ and $\theta_{k,l}$, respectively. 

    \subsection{Digital Twin Model}
        Wireless digital twins are virtual representations of the physical communication environment, which can be used to generate synthetic channel information~\cite{Alkhateeb2023}. In the real world, wireless channels are determined by three factors: (i) Communication environment $\cE$ comprises the positions, orientations, dynamics, and shapes of the BS, users, and surrounding objects, e.g., reflectors and scatterers. (ii) Signal propagation law g(.) determines how signals propagate through the environment. (iii) Hardware characteristics $\cS$ specify the physical properties and impairments of communication hardware. Given these components, the channel set for the $K$ users $\boldsymbol{\cH} = \{\bh_1, \ldots, \bh_K\}$ can be expressed as:
        \begin{equation}
            \boldsymbol{\cH} = g(\cE, \cS).
        \end{equation}
        Digital twins approximate the communication environment using a 3D EM model and simulate signal propagation via ray tracing. Hardware characteristics can be modeled through a combination of measurement and analytical techniques. Thus, the synthetic channel $\widetilde{\boldsymbol{\cH}} = \{\widetilde{\bh}_1, \ldots, \widetilde{\bh}_K\}$ generated by the site-specific digital twin can be written as
        \begin{equation}
            \widetilde{\boldsymbol{\cH}} = \widetilde{g}(\widetilde{\cE}, \widetilde{\cS}),
        \end{equation}
        where $\widetilde{g}(.)$, $\widetilde{\cE}$, and $\widetilde{\cS}$ denote the ray tracing algorithm, the 3D EM model, and the hardware model, respectively.

\section{Problem Formulation}
    In this work, we aim to refine the synthetic channel information generated by a low-complexity digital twin using a lightweight deep learning model. We specifically focus on refining the DFT-domain channel information, which is represented by the DFT weights of the user channel. The DFT weights for the $k^{\text{th}}$ user are computed as $\bz_k = | \bD^H \bh_k | \in \bbR^{N \times 1}$, where $\bD = [\bd_1, \ldots, \bd_N] \in \bbC^{N \times N}$ is the DFT matrix comprising $N$ orthogonal DFT vectors. When we apply the DFT matrix to the spatial-domain channel, it can be viewed as projecting the channel onto a set of orthogonal beams, i.e., DFT beams. Thus, the DFT weights can also be interpreted as the correlation between the user channel and each DFT beam that covers a specific angular direction. Aside from the DFT weights, the user's position can also provide useful information for refining the channel information. This is because the user's position helps to resolve ambiguities in the DFT weights, especially when the user is in a non-line-of-sight (NLoS) region. Similar DFT weights may require different refinements according to the geometry around the users. Therefore, we consider both the user position and the DFT weights generated by the digital twin as inputs to our calibration model.

    The goal of the calibration is to learn a mapping function to refine the DFT weights generated by the low-complexity digital twin. Let $\bp_k \in \bbR^{3 \times 1}$ denote the 3D position of the $k^{\text{th}}$ user, and $\widetilde{\bz}_k = | \bD^H \widetilde{\bh}_k | \in \bbR^{N \times 1}$ represent the DFT weights obtained from the low-complexity digital twin. The mapping function can be denoted as $f(\bp_k, \widetilde{\bz}_k; \Theta)$, where $\Theta$ is the set of learnable parameters. The optimization problem can be formulated as
    \begin{gather}  \label{eq:opt_calibration}
        \begin{align}
            \min_{\Theta} \quad & \sum_{k=1}^{K} \| \bz_k^\star - \widehat{\bz}_k \|_2 \nonumber \\
            \textrm{s.t.} \quad & \widehat{\bz}_k = f(\bp_k, \widetilde{\bz}_k; \Theta),\ \forall k,
        \end{align}
    \end{gather}
    where $\bz_k^\star$ is the ground truth DFT weights, and $\widehat{\bz}_k$ is the refined DFT weights predicted by the mapping function. The objective function in \eqref{eq:opt_calibration} minimizes the difference between the refined DFT weights and the ground truth DFT weights for all users. In the next section, we present the proposed training approaches for obtaining the ground truth DFT weights and the model design for the mapping function.

\section{Proposed Solution}
    In this section, we present the proposed digital twin calibration approach to refine the DFT-domain channel information generated by a low-complexity digital twin. We also discuss its application in codebook-based CSI feedback.
    \subsection{Key Idea}
        A wireless digital twin with higher fidelity can provide more accurate synthetic channel information; however, it often comes with higher computational costs, which makes it unsuitable for real-time applications. In contrast, a low-complexity digital twin can generate synthetic channel information more quickly but may compromise accuracy. To address this trade-off, we propose a calibration approach that refines the DFT-domain channel information generated by a low-complexity digital twin. The calibration is performed using a lightweight deep learning model, which offers the following two benefits:
        \begin{itemize}
            \item \textbf{Fidelity enhancement}: A high-fidelity digital twin can provide more accurate synthetic channel information, but its computational demands may be prohibitive for real-time applications. The proposed calibration approach can learn from the high-fidelity digital twin in an offline phase, offering a faster alternative to directly using a high-fidelity digital twin in the online phase.
            \item \textbf{Real-world adaptation}: DFT-based codebooks are widely used in current standardized wireless systems~\cite{3GPP}, and it is feasible to collect historical CSI feedback data from these systems. With this data, we can train the calibration model to adapt the digital twin to the real-world communication environment, bridging the gap between synthetic and actual channel information.
        \end{itemize}
        In the following subsections, we present the details of the proposed digital twin calibration approach.

    \subsection{Training Approaches of the Calibration Model}
        In this subsection, we introduce two training approaches for the calibration model. This calibration is achieved using deep learning with supervised training. Labeled data for training can be generated in two ways, depending on the target context:
        \begin{itemize}
            \item \textbf{High-fidelity digital twin}: When a high-fidelity digital twin is the refining target, we can directly obtain the true DFT weights from its synthetic channel information. In an offline phase, we generate a paired dataset for training that includes user positions and the corresponding DFT weights from both the low-complexity and high-fidelity digital twins. This dataset is then used to train our model to refine the output of the low-complexity twin.
            \item \textbf{Real world}: When the goal is to adapt the digital twin to the real world, historical CSI feedback from current standardized wireless systems can serve as labels. For example, 3GPP Type-II codebook allows users to report the best 2 to 4 DFT beam indices along with their corresponding coefficients~\cite{3GPP}. By correlating user positions with historical CSI feedback, we can construct labeled datasets necessary for training the calibration model.
        \end{itemize}
        In this work, we adopt the U-Net architecture~\cite{Ronneberger2015} to learn the calibration. The user's position is first passed through an embedding layer to obtain a fixed-size vector representation, which is then concatenated with the DFT weights. The calibration process can be expressed as
        \begin{equation}
            \bee_k = f_{\rm{embed}}(\bp_k),
        \end{equation}
        \begin{equation}
            \widehat{\bz}_k = f_{\rm{UNet}}(\bee_k, \widetilde{\bz}_k),
        \end{equation}
        where $f_{\rm{embed}}(.)$ is the embedding function that maps the user position to a fixed-size vector, and $f_{\rm{UNet}}(.)$ is the U-Net model that learns the mapping from the user position and the DFT weights to the refined DFT weights. The loss function for training the calibration model can be defined as the mean squared error (MSE), given by
        \begin{equation}
            \cL(\bz_k^\star, \widehat{\bz}_k) = \| \bz_k^\star - \widehat{\bz}_k \|_2^2.
        \end{equation}
        In the first training approach, $\bz_k^\star$ is obtained from the high-fidelity digital twin, allowing us to use the full DFT weights as ground truth. For the second training approach, $\bz_k^\star$ is derived from historical CSI feedback, which provides the top $P$ DFT beam indices and their corresponding coefficients. Consequently, $\bz_k^\star$ is subject to the constraint that only $P$ codewords are selected, i.e., $\| \bz_k^\star \|_0 = P$.

    \begin{table}[t]
        \centering
        \caption{Proposed U-Net Model Architecture}
        \label{table:unet_architecture}
        \begin{threeparttable}
        \footnotesize
        \setlength{\tabcolsep}{1.1pt}
        \begin{tabular}{llcccc}
            \toprule
            \textbf{Layer} & \textbf{Module} & \textbf{K.} & \textbf{S.} & \textbf{Input Shape} & \textbf{Output Shape} \\
            \midrule
            \multicolumn{6}{l}{\textbf{Encoder}} \\
            \midrule
            Input \& Concat. & MLP & - & - & $(B, N) \text{ \& } (B, 3)$ & $(B, 2, N)$ \\
            Layer 1 & Conv. Block & 3 & 1 & $(B, 2, N)$ & $(B, 16, N)$ \\
            Downsample & MaxPool & 3 & 2 & $(B, 16, N)$ & $(B, 16, N/2)$ \\
            Layer 2 & 2$\times$ Conv. Blocks & 3 & 1 & $(B, 16, N/2)$ & $(B, 32, N/2)$ \\
            Downsample & MaxPool & 3 & 2 & $(B, 32, N/2)$ & $(B, 32, N/4)$ \\
            Layer 3 & 2$\times$ Conv. Blocks & 3 & 1 & $(B, 32, N/4)$ & $(B, 64, N/4)$ \\
            Downsample & MaxPool & 3 & 2 & $(B, 64, N/4)$ & $(B, 64, N/8)$ \\
            Layer 4 & 2$\times$ Conv. Blocks & 3 & 1 & $(B, 64, N/8)$ & $(B, 128, N/8)$ \\
            \midrule
            \multicolumn{6}{l}{\textbf{Decoder}} \\
            \midrule
            Upconv & ConvTranspose & 3 & 2 & $(B, 128, N/8)$ & $(B, 64, N/4)$ \\
            Layer 1 & 2$\times$ Conv. Blocks & 3 & 1 & $(B, 64 + 64, N/4)$ & $(B, 64, N/4)$ \\
            Upconv & ConvTranspose & 3 & 2 & $(B, 64, N/4)$ & $(B, 32, N/2)$ \\
            Layer 2 & 2$\times$ Conv. Blocks & 3 & 1 & $(B, 32 + 32, N/2)$ & $(B, 32, N/2)$ \\
            Upconv & ConvTranspose & 3 & 2 & $(B, 32, N/2)$ & $(B, 16, N)$ \\
            Layer 3 & 2$\times$ Conv. Blocks & 3 & 1 & $(B, 16 + 16, N)$ & $(B, 16, N)$ \\
            \midrule
            \multicolumn{6}{l}{\textbf{Output}} \\
            \midrule
            Output Layer & 1$\times$ Conv. Block & 3 & 1 & $(B, 16, N)$ & $(B, 1, N)$ \\
            Final Output & Softmax & - & - & $(B, 1, N)$ & $(B, 1, N)$ \\
            \bottomrule
        \end{tabular}
        \begin{tablenotes}
        \footnotesize
        \item Note: \textbf{K.} and \textbf{S.} denote the kernel size and stride, respectively. $B$ denotes the batch size. MLP stands for multi-layer perceptron.
        \end{tablenotes}
        \end{threeparttable}
    \end{table}

    \subsection{Deep Learning Model Architecture}
        We adopt a 1-dimensional U-Net architecture for the calibration model, as summarized in \tabref{table:unet_architecture}. The model takes two inputs: DFT weights and a $N$-dimensional embedded representation of the user's position. The position is processed by a multi-layer perceptron to provide contextual information. The U-Net follows an encoder-decoder structure with skip connections. The encoder down-samples the feature maps through three layers, each consisting of two sequential sets of convolutional, batch normalization, and ReLU layers, followed by a max-pooling operation. The decoder mirrors this, up-sampling the features with transpose convolutions. Skip connections concatenate feature maps from corresponding encoder layers to their upsampled counterparts in the decoder. The final output layer is a convolutional operation followed by a softmax activation, which produces the refined DFT coefficients.

    \subsection{Case Study: Codebook-Based CSI Feedback}
        Now that we have established the digital twin calibration approach, an important question remains: How can we utilize the refined DFT-domain channel information to assist wireless communication tasks? To answer this question, we consider codebook-based CSI feedback as a case study, which is illustrated in \figref{fig:system_model}. The existing codebook-based CSI feedback methods, e.g., Type-I/II codebooks in 3GPP~\cite{3GPP}, require full channel estimation at the user, which incurs high pilot signal overhead, particularly for large-scale MIMO systems. However, with the assistance of digital twins, we can leverage the synthetic channel information to reduce this overhead. Specifically, we use the refined DFT-domain channel information to identify the most relevant DFT beams for each user before the CSI feedback process. The digital twin aided CSI feedback process consists of the following steps. First, the BS generates synthetic channel information $\widetilde{\bz}_k$ using digital twin. Then, the synthetic channel information is refined using the trained calibration model, resulting in $\widehat{\bz}_k$. The refined channel information serves as prior knowledge for the BS to identify the most relevant codewords for each user. The BS selects the $P$ codewords that have the highest values in $\widehat{\bz}_k$, which can be expressed as
        \begin{equation}
            \{ \, \widehat{i}_{k, 1}, \ldots, \widehat{i}_{k, P}\} = \argsort_{i=\{1, \ldots, N\}} \left[ \widehat{\bz}_k \right]_{i}.
        \end{equation}
        The BS then transmits pilot signals precoded by these selected DFT codewords, allowing users to estimate their coefficients. The received signal at user $k$ can be expressed as
        \begin{equation}
            \widehat{\by}_k = \bh_k^H \widehat{\bQ}_k  \bS_k + \bn_k,
        \end{equation}
        where $\widehat{\bQ}_k \in \bbC^{N \times P}$ is the sub-matrix of the DFT codebook corresponding to the selected codewords. $\bS_k \in \bbC^{P \times P}$ is the matrix that contains $P$ transmitted pilot symbols on its diagonal. $\bn_k \sim \cN(0, \sigma^2 \bI)$ is the noise vector at the $k^{\text{th}}$ user. This measurement process can be viewed as projecting the channel onto the selected DFT codewords, which enables users to estimate the effective channel coefficients. Based on the received signal, each user can estimate the effective channel coefficients and report the normalized coefficients back to the BS, denoted as $\widehat{\bx}_k \in \bbC^{P \times 1}$. Finally, the BS reconstructs the CSI from the reported coefficients as a linear combination of the selected codewords, i.e., $\widehat{\bw}_k = \widehat{\bQ}_k \widehat{\bx}_k$, and designs the precoding vector $\widehat{\bff}_k$ accordingly. This step is similar to existing codebook-based CSI feedback methods. Overall, this digital twin aided approach leverages refined synthetic channel information to reduce downlink pilot transmission overhead while maintaining the same feedback overhead as existing methods. Notably, this approach is not limited to codebook-based CSI feedback but can also be applied to beam selection for mmWave frequency bands~\cite{Giordani2018}.

        \begin{figure}[t]
            \centering
            \includegraphics[width=0.475\textwidth]{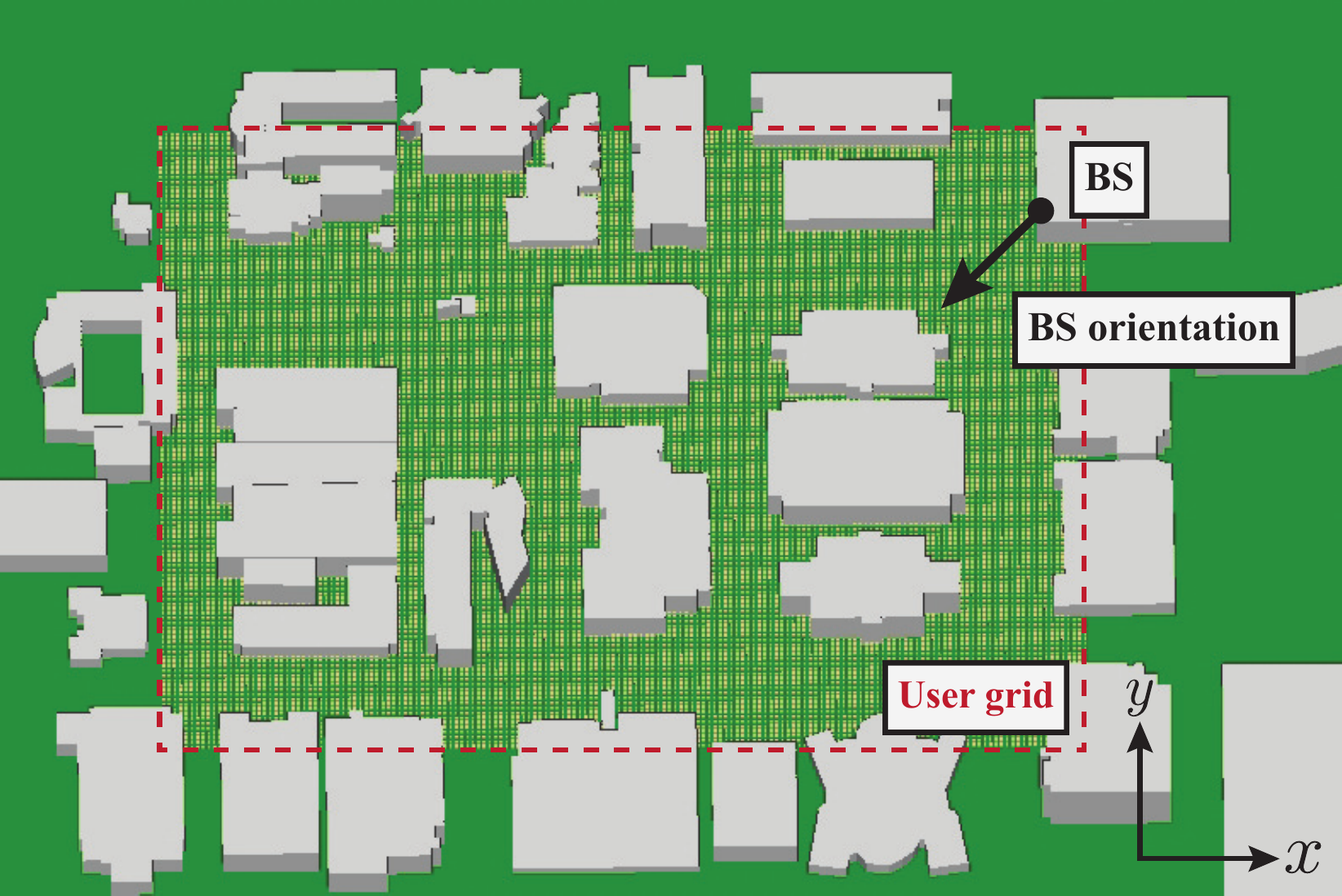}
            \caption{This figure shows the bird's-eye view of the Arizona State University (ASU) campus, which serves as the study area for the simulation. The BS is located at the rooftop of a building at the top right corner, and the user grid is highlighted by the red box.}
            \label{fig:asu}
        \end{figure}

        \begin{figure*}[t]
            \centering
            \includegraphics[width=0.975\textwidth]{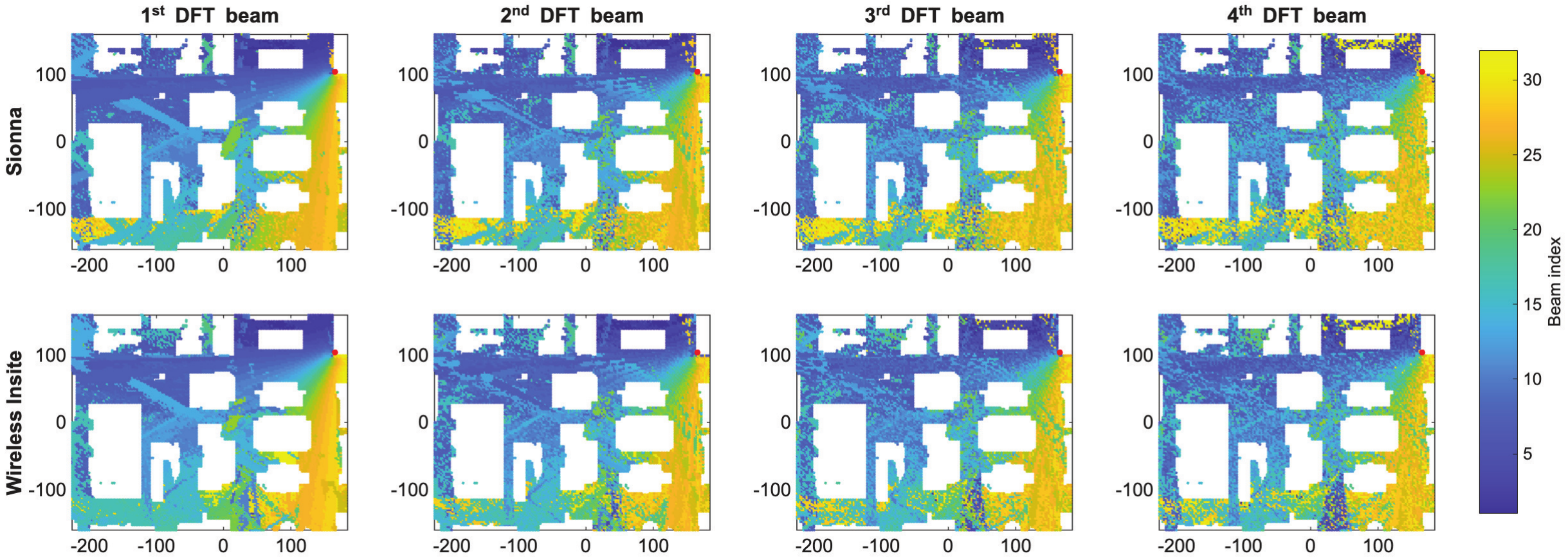}
            \caption{This figure presents the heatmaps illustrating the top-4 DFT beam indices selected by the Sionna RT (baseline) and Wireless Insite (target) ray tracers. The key observation is that larger differences occur when the user is situated in the NLoS region (the upper and lower-left areas of the user grid), which is aligned with the fact that the baseline scenario employs a ray tracer with a simplified diffraction model.}
            \label{fig:beam_idx_difference}
        \end{figure*}

\section{Simulation Results}
    \subsection{Simulation Setup}
        \textbf{Scenario setup:} In the simulation, we consider two ray-tracing scenarios. First, the \textit{target scenario} represents either a high-fidelity digital twin or a real-world communication environment, with its geometry reflecting the Arizona State University (ASU) campus, as illustrated in~\figref{fig:asu}. For this scenario, we used Wireless Insite~\cite{Remcom}, a high-complexity ray tracer, to simulate signal propagation. We set the maximum number of reflections and diffractions to $6$ and $1$, respectively, and enabled diffuse scattering. Second, the \textit{baseline scenario} serves a low-fidelity digital twin, utilizing Sionna RT~\cite{Hoydis2023} for ray tracing. In this scenario, we set the maximum number of interactions to 6, with reflections, diffractions, and scattering all enabled. It is important to note that Sionna v0.19.2 assumes diffraction is the only interaction if it exists in a path. While this significantly reduces the computational complexity of the ray-tracing simulation, it can result in less accurate channel information. For both scenarios, we consider a BS equipped with a uniform linear array (ULA) of $N=32$ antennas, positioned at a height of $22$ m. The operation frequency is set to $3.5$ GHz. A user grid is deployed near the BS, measuring $410$ m $\times$ $320$ m, with single-antenna users uniformly distributed at a spacing of $2.5$ m. Finally, a DFT codebook comprising $32$ codewords is adopted, and the number of selected codewords for each user's CSI feedback is set to $P=4$.

        \textbf{Dataset generation:} We generate our dataset by performing ray-tracing simulations in both scenarios. This process yields essential path parameters, including the complex gain $\alpha_{k,l}$, as well as the azimuth and elevation AoDs, $\phi_{k,l}$ and $\theta_{k,l}$. Subsequently, we employ the DeepMIMO channel generator~\cite{Alkhateeb2019} to produce the user channels. The channels of both scenarios are then utilized to construct the paired data necessary for training our calibration model. The training dataset comprises $13191$ samples. To evaluate the spatial generalization ability of the calibration model, we also generate an additional $7895$ off-the-grid samples. These samples are distinct from the training dataset, created by randomly selecting user positions and simulating their channel coefficients using the ray tracers.       

        \begin{figure}[t]
            \centering
            \includegraphics[width=0.475\textwidth]{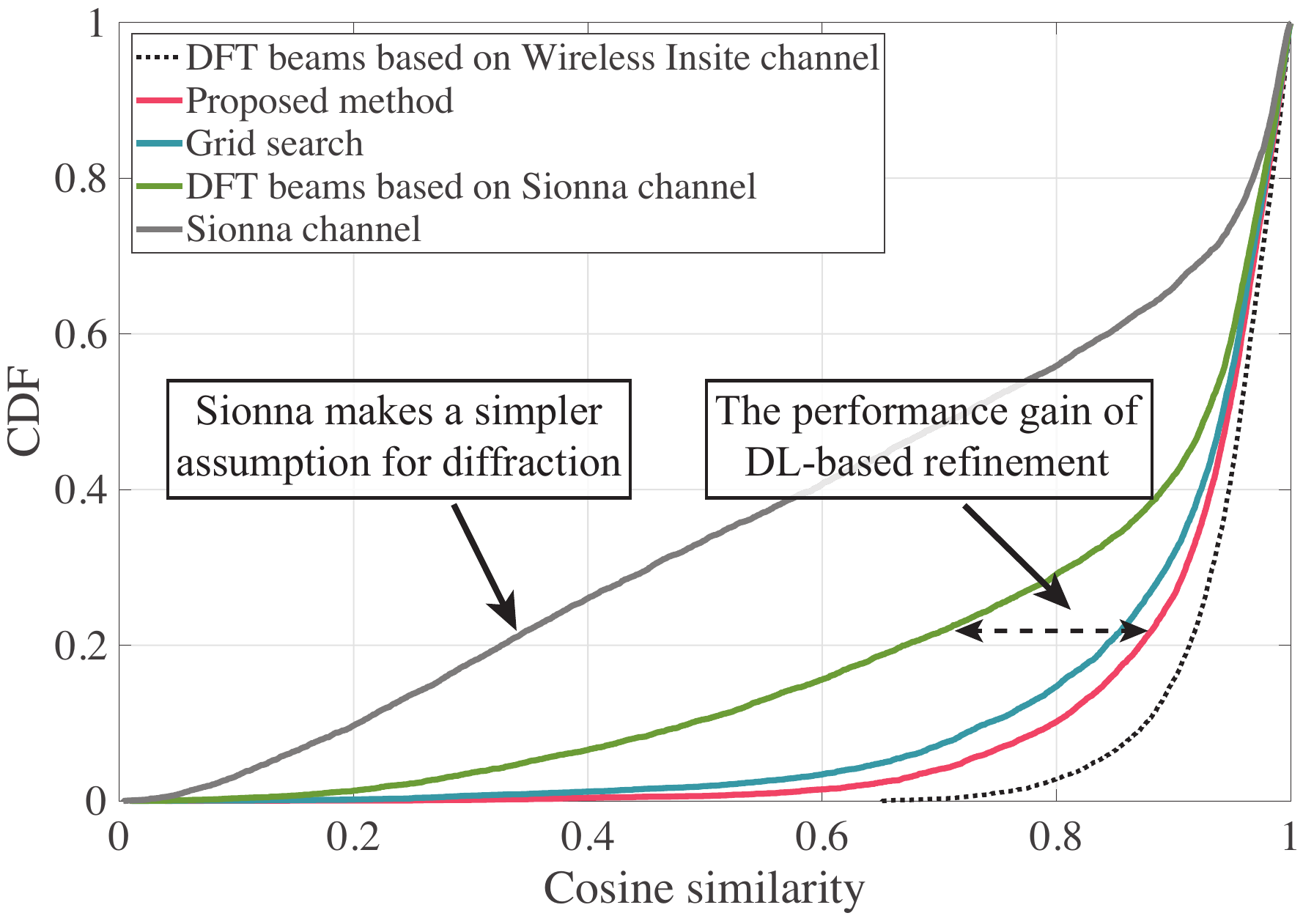}
            \caption{This figure presents the CDF of the cosine similarity between the ground-truth channel and the estimated CSI using the top-4 DFT beams selected by the proposed calibration approach and the benchmark scenarios. By integrating a lightweight calibration model (inference: $0.0018$ s/sample), our approach nears the performance upper bound while maintaining a total cost of $0.0610$ s/sample, which is significantly lower than the $1.2019$ s/sample required by a high-fidelity digital twin.}
            \label{fig:cdf_cos_sim}
        \end{figure}

    \subsection{Performance Evaluation}
        In this subsection, we evaluate the performance of the proposed digital twin calibration approach. The main performance metric is the cosine similarity between the ground-truth channel and the estimated CSI, defined as
        \begin{equation}
            \rho(\bh, \widehat{\bw}) = \frac{|\bh^H \, \widehat{\bw}|}{\|\bh\|_2 \|\widehat{\bw}\|_2}.
        \end{equation}
        Cosine similarity measures the directional alignment between the actual channel and the estimated CSI; higher values indicate more accurate CSI and greater potential for precoding performance. Next, we address the following key questions:

        \textbf{What is the impact of a low-complexity digital twin on the performance?} We first address this question by comparing the top-4 DFT beam indices selected in the target and baseline scenarios. In \figref{fig:beam_idx_difference}, we present heatmaps showing the absolute differences between these top-4 DFT beam indices. The main observation is that larger differences occur when the user is in the NLoS region, which is consistent with the fact that the baseline scenario uses a ray tracer with a simplified diffraction model. Another perspective is to evaluate the performance when the channel from the low-complexity digital twin is used directly as the estimated CSI, i.e., $\widehat{\bw}_k = \widetilde{\bh}_k$. This serves as a benchmark to assess the impact of relying solely on a low-complexity digital twin. In \figref{fig:cdf_cos_sim}, we include this benchmark and show its performance in the cumulative distribution function (CDF) of cosine similarity. The results indicate that directly using the channel from the low-complexity digital twin leads to poor performance, highlighting the necessity of a calibration approach to refine this channel information.

        \textbf{How effective is the proposed calibration approach?} To evaluate the effectiveness of our proposed calibration approach, we compute the cosine similarity for the estimated CSI obtained using the top-4 DFT beams selected by our method. For comparison, we compute the cosine similarity for the target and baseline scenarios, where the estimated CSI is derived from the top-4 DFT beams selected by their respective scenarios. These serve as upper and lower bounds for performance. We also compare our proposed method against a grid search benchmark. Given the user position for evaluation, the grid search method identifies the nearest neighbor in the training user grid of the target scenario and uses the corresponding beams from that position. In \figref{fig:cdf_cos_sim}, the results indicates that the proposed calibration approach achieves significantly higher cosine similarity compared to the baseline scenario. Also, the proposed method consistently outperforms the grid search benchmark. This shows that the calibration model can effectively leverage both user position and low-fidelity channel information to generalize to unseen user positions. Furthermore, the performance of our proposed approach is close to that of the target scenario, demonstrating its effectiveness in refining low-fidelity channel information.

        \textbf{How much computational overhead can be reduced?} To assess the computational efficiency, we analyze the computation time required for the steps involved in the process, which is conducted on an Nvidia RTX A5000 GPU. This includes the time taken for ray tracing, as well as the time required for the refinement. The findings indicate that the computation time for ray tracing using Sionna RT (0.0592 seconds per sample) is much lower than that of Wireless Insite (1.2019 seconds per sample). Also, the number of parameters in our U-Net model is only $181$K, leading to a lightweight model where the inference time is minimal (0.0018 seconds per sample). Overall, the total computation time for the combination of Sionna RT and DL-based refinement is substantially lower than that of Wireless Insite alone. This shows our proposed calibration approach can effectively leverage a low-complexity digital twin to achieve performance comparable to a high-fidelity digital twin, while significantly reducing computational overhead.

\section{Conclusion}
    In this paper, we study the calibration problem of digital twins, which aims to refine the synthetic channel information generated by a low-complexity digital twin using a lightweight deep learning model. We present two training approaches, which can learn from either a high-fidelity digital twin or historical CSI feedback from standardized wireless systems. We demonstrate the effectiveness of the proposed approach with a case study of codebook-based CSI feedback, where the refined synthetic channel information is used to identify the most relevant codewords for each user. Simulation results highlight the effectiveness of our proposed approach in improving CSI acquisition accuracy. For future work, an interesting direction is to extend this framework to support adapting the digital twin to dynamic environments based on limited feedback, such as received power \cite{Alikhani2025}. This would enable the digital twin to continuously learn and adapt to changing conditions, further enhancing its utility in real-world applications.

\section*{Acknowledgement} This work was supported in part by the National Science Foundation under Grant No. 2426906.


\end{document}

%% file: input.tex
\usepackage{amsfonts}
\usepackage{times}
\usepackage{graphicx}
\usepackage{latexsym}
\usepackage{dsfont}
\usepackage{amssymb}
\usepackage{amsmath}
\usepackage{cite}
\usepackage{verbatim}

\newcommand{\figref}[1]{{Fig.}~\ref{#1}}
\newcommand{\tabref}[1]{{Table}~\ref{#1}}

\def\bb0{{\mathbb{0}}}

\def\ba{{\mathbf{a}}}
\def\bb{{\mathbf{b}}}

\def\bd{{\mathbf{d}}}
\def\bee{{\mathbf{e}}}
\def\bff{{\mathbf{f}}}

\def\bh{{\mathbf{h}}}

\def\bn{{\mathbf{n}}}

\def\bp{{\mathbf{p}}}

\def\bw{{\mathbf{w}}}
\def\bx{{\mathbf{x}}}
\def\by{{\mathbf{y}}}
\def\bz{{\mathbf{z}}}
\def\b0{{\mathbf{0}}}

\def\bD{{\mathbf{D}}}

\def\bI{{\mathbf{I}}}

\def\bQ{{\mathbf{Q}}}

\def\bS{{\mathbf{S}}}

\def\bbC{{\mathbb{C}}}

\def\bbR{{\mathbb{R}}}

\def\cE{\mathcal{E}}

\def\cH{\mathcal{H}}

\def\cL{\mathcal{L}}

\def\cN{\mathcal{N}}

\def\cS{\mathcal{S}}

\def\sf0{{\mathsf{0}}}

